\documentclass[letter,traditabstract]{aa}
\usepackage{graphicx}
\usepackage{times,amssymb,amsmath} 
\usepackage{epsfig} 
\usepackage{lscape}
\usepackage[usenames]{color}
\usepackage{txfonts}

\begin{document}
  \title{Ly$\alpha$ emitters: blue dwarfs or supermassive ULIRGs? \\
Evidence for a transition with redshift}

\titlerunning{Ly$\alpha$ emitting ULIRGs}
 
  \author{K.K. Nilsson
         \inst{1}
         \and
        P. M{\o}ller\inst{2}
         }

  \institute{ST-ECF, Karl-Schwarzschild-Stra\ss e 2, 85748, Garching bei M\"unchen, Germany\\
                   \email{knilsson@eso.org} 
                   \and
                   European Southern Observatory, Karl-Schwarzschild-Stra\ss e 2, 85748 
                   Garching bei M\"unchen, Germany}

  \date{Received date; accepted date}

\abstract
{The traditional view that Ly$\alpha$ emission and dust should be mutually exclusive
has been questioned more and more often; most notably, the observations of
Ly$\alpha$ emission from ULIRGs seem to counter this view. In this
paper we seek to address the reverse question. How large a fraction of
Ly$\alpha$ selected galaxies are ULIRGs? Using two samples of 24/25 Ly$\alpha$ emitting galaxies at $z = 0.3/2.3$, we perform this test, including results at $z = 3.1$, and find that, whereas the
ULIRG fraction at $z = 3.1$ is very small, it systematically increases
towards lower redshifts. There is a hint that this evolution may be
quite sudden and that it happens around a redshift of $z \sim 2.5$. After measuring the infrared luminosities of the Ly$\alpha$ emitters, we find that they are in the normal to ULIRG range in the lower redshift sample, while the higher redshift galaxies all have luminosities in the ULIRG category. The Ly$\alpha$ escape fractions for these infrared bright galaxies are in the range $1-100$\% in the $z = 0.3$ galaxies, but are very low in the $z = 2.3$ galaxies, 0.4\% on average. The unobscured star formation rates are very high, ranging from 500 to more than 5000 M$_{\odot}$~yr$^{-1}$, and the dust attenuation derived are in the range $0.0 < A_V < 3.5$. }


  \maketitle

\section{Introduction}
Galaxies at high redshift come in all kinds of flavours. Depending on the search criteria, the sources found may be dusty or dust-free, more or less massive, star forming or quiescent. Some of the most impressive beasts of the high redshift zoo are the sub-mm and ultra-luminous infrared galaxies: SMGs (Ivison et al. 1998, 2000, Blain et al. 2002, Chapman et al. 2005) and ULIRGs (Sanders \& Mirabel 1996). These galaxies show star formation rates in the hundreds or thousands of solar masses per year, with most of their light re-processed into the infrared or sub-mm wavelengths due to large amounts of dust. On the other end of the scale, Ly$\alpha$ emitters are found (LAEs; M{\o}ller \& Warren 1998, Gronwall et al. 2007,
Nilsson et al. 2007, Finkelstein et al. 2009, Ouchi et al. 2008, Grove et al. 2009) which have hitherto been generally considered blue and dust-free galaxies with moderate star formation rates of a few to a few times ten solar masses per year. 

Only a few publications have reported finding red and/or dusty Ly$\alpha$ emitters (Stiavelli et al.~2001, Colbert et al.~2006, Lai et al.~2007, Finkelstein et al.~2009, Nilsson et al.~2009). Considering the traditional view that even small amounts of dust would quench any Ly$\alpha$ emission (for a discussion, see Pritchet 1994), the detection of the Ly$\alpha$ line in dusty galaxies would seem surprising, but even more so the
detection of Ly$\alpha$ in a sample of sub-mm galaxies (Chapman et al.~2003, 2005). Likely explanations for the existence of dusty galaxies with Ly$\alpha$ emission are either special geometrical alignments or strong local variations in the dust-to-gas ratios in a clumpy medium (Neufeld 1991, Hansen \& Oh 2006). Observational evidence that Ly$\alpha$
emission is commonly seen among sub-mm galaxies does not, however, constitute evidence for the opposite,
since sub-mm galaxies are rare and because there are few ways other than Ly$\alpha$ to determine the redshift of very red galaxies. It follows
that the overlap reported so far could well be an observational
selection bias.
The fundamental questions that remain are therefore whether {\it i)}
sub-mm galaxies and ULIRGs are common among Ly$\alpha$ emitters and
{\it ii)} whether the ULIRG fraction evolves with redshift. This \emph{Letter} aims to address
those two questions. We here present the infrared properties of two samples of galaxies found through their Ly$\alpha$ emission at $z = 0.3$~and~$2.3$. We show that they have infrared fluxes well into the ULIRG regime. This sample is unique, bridging the gap between low-mass star forming and large star-bursting galaxies.

Throughout this paper, we assume a cosmology with $H_0=72$
km s$^{-1}$ Mpc$^{-1}$, $\Omega _{\rm m}=0.3$ and
$\Omega _\Lambda=0.7$.

\section{Data}
The Ly$\alpha$ candidate samples studied here come from Deharveng et al. (2008, $z=0.3$, hereafter z03) and Nilsson et al.~(2009, $z=2.3$, hereafter z23). The z03 sample consists of a total of 31 Ly$\alpha$ emitting galaxies at $0.1 < z < 0.4$ in the ECDF-S and ELAIS-S fields, found in a survey volume of $\sim 5 \times 10^5$~Mpc$^3$. Of the 31, one is observed to be an AGN by Cowie et al.~(2009) and 11 are not included in their classification. At $z = 2.3$, the sample includes 187 Ly$\alpha$ emitting candidates at redshifts $2.21< z <2.31$, and are spread over a $\sim 0.2$~deg$^2$ area in the central COSMOS field (the survey volume is $\sim 3\times10^5$~Mpc$^3$). Of the 187 candidates, 27 are considered AGN, based on detections in public \emph{Chandra} and/or XMM X-ray images. The ECDF-S, ELAIS-S and COSMOS fields are covered by the SWIRE (Lonsdale et al.~2003) and S-COSMOS (Sanders et al.~2007) \emph{Spitzer} surveys. 
The photometry of all the z23 candidates, in all \emph{Spitzer} bands from $3.6 - 24\mu$m, will be presented in a forthcoming publication (Nilsson et al., in prep.). 
Here we focus on the results found for the photometry in the $8\mu$m and $24\mu$m (MIPS) bands for the z03 and z23 samples, respectively, corresponding to $\sim 7 \mu$m in the restframe of both samples. 

For the z03 emitters, aperture photometry centred on the
Ly$\alpha$ coordinates was performed. The aperture radius was $4.5''$,
selected to be consistent with the z23 results. The limiting
sensitivity is a few times ten $\mu$Jy. Of the 31
candidates, 24 are detected at $> 3\sigma$, of which
Cowie et al.~(2009) classified one as an AGN and 17 as galaxies.

In COSMOS, both a deep and a wide survey has been published in the MIPS band. The deep survey covers $\sim 33$\% of the field surveyed for Ly$\alpha$, and the sensitivities reached in the two surveys were 71 and 150 $\mu$Jy, respectively (5$\sigma$). To find counterparts to the Ly$\alpha$ emitters, the public catalogues were searched within 4 pixels ($4.8''$, i.e. $0.8 \, \times$~ FWHM of the MIPS PSF) radii of each source and 25 counterparts were found. Of these, 16 are also X-ray detected, and are considered to be AGN based on their R band-to-X-ray flux ratios. The
fluxes of the galaxies in the 8/24~$\mu$m bands are found in
Table~\ref{tab:IR}.

\section{Infrared properties of Ly$\alpha$ emitters}
\begin{figure}[!t]
\begin{center}
\epsfig{file=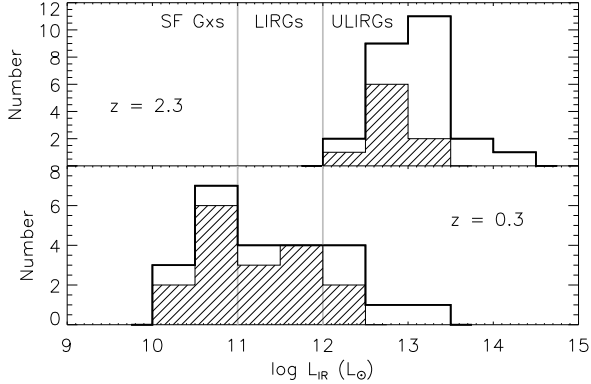,width=8.0cm}
\caption{Histogram of infrared luminosities of all IR detected sources as derived with Eq.~\ref{eq:IR}.
At both redshifts, the hatched area marks genuinely certain
LAE galaxies, while the open areas are AGN or unclassified
objects. 
Solid vertical lines mark the LIRG/ULIRG definitions.}
\label{fig:IRnum}
\end{center}
\vspace{-0.6cm}
\end{figure}

At $z = 2.25$, the MIPS $24\mu$m band corresponds to restframe $5.9 - 8.8$~$\mu$m. Correspondingly, the $8\mu$m IRAC band covers $4 - 8.8$~$\mu$m for the z03 emitters. To convert this mid-infrared luminosity to the total infrared luminosity, we use the conversion of Chary \& Elbaz (2001): 
\begin{equation}\label{eq:IR}
L_{IR} \, [8-1000 \, \mu\mathrm{m}] = 4.37^{+2.35}_{-2.13} \times 10^{-6}L^{1.62}_{6.7 \mu m}.
\end{equation}
The values derived can be found in Table~\ref{tab:IR}, and a histogram of the total infrared luminosities is shown in Fig.~\ref{fig:IRnum}.
\begin{table*}[!t]
\begin{center}
\caption{IR fluxes, star formation rates and derived dust attenuation. }
\begin{tabular}{@{}lcccccccc}
\hline
\hline
\multicolumn{8}{c}{$z = 2.3$} \\
\hline
LAE\_ & $F_{24\mu m}$ & $L_{IR}$ &  $L_{bol}$ & $SFR_{bol}$ & $A_{1600}$ & Offset & Remark \\
COSMOS\_\# & $\mu$Jy & erg s$^{-1}$ & erg s$^{-1}$ & M$_{\odot}$ yr$^{-1}$ & & arcsec &  &  \\
\hline
  20 & $154\pm16$ &  $46.50\pm0.05$ &  $46.50\pm0.07$ & $     1420\pm220$   & $5.51\pm0.18$ & 0.02 & AGN \\
  36 & $835\pm20$ &  $47.69\pm0.02$ &  $47.69\pm0.02$ & $   21830\pm1000$ & $7.65\pm0.05$ & 0.02 & AGN \\
\hline
\multicolumn{8}{c}{z = 0.3} \\
\hline
ID & $F_{8\mu m}$   &  $L_{IR}$ &  $L_{bol}$ & $SFR_{bol}$ & $A_{1600}$ & --- & Remark & \\
    & $\mu$Jy & erg s$^{-1}$ & erg s$^{-1}$ & M$_{\odot}$ yr$^{-1}$ &  &  \\
\hline
CDFS\_1348 & $1.63\pm0.03$   & $45.19\pm0.01$ & $45.20\pm0.01$ & $70.9\pm2.1$   & $4.91\pm0.03$ &  & \\
CDFS\_1821 & $0.38\pm0.05$   & $44.17\pm0.08$ & $44.20\pm0.09$ &$7.1\pm1.4$      & $2.69\pm0.21$ & &   Unclass. & \\
\hline
\label{tab:IR}
\end{tabular}
\end{center}
\begin{list}{}{}{}{}{}{} 
\item[ ] Columns are: (1) flux in the 8/24~$\mu$m data, (2) infrared luminosity, (3) bolometric luminosity, (4) star formation rate, (5) dust attenuation derived from the infrared luminosities, (6) offsets between MIPS catalogue source and LAE candidate in z23 sample, and (7) other remarks. The IDs for the z03 sample are from Deharveng et al.~(2008). Full table is available in the online version.
\end{list}
\end{table*}
At each redshift, the infrared
luminosities for the full samples (25 and 24, respectively), as well as the subsamples with certain identifications, are shown. 
 It is seen that the flux limits at the two redshifts cause the overlap between the high and z03 sample to be very small.
All of the z23 sources are consistent with ULIRG luminosities ($L_{IR} > 10^{12} L_{\odot}$), including the ``normal'' LAEs. The luminosities are also in the same range as high redshift sub-mm galaxies (Chapman et al. 2005). At $z = 0.3$, roughly half of the galaxies lie in the range of normal star forming galaxies, seven have LIRG luminosities ($10^{11} < L_{IR} < 10^{12} L_{\odot}$), and two have ULIRG luminosities. Note that at $z = 2.3$, detected sources are automatically ULIRGs, as the detection limit in the deep survey in the $24\mu$m band corresponds to $\log L_{IR} = 12.4 \, L_{\odot}$, and in the shallow survey to $\log L_{IR} = 12.9 \, L_{\odot}$.

\begin{figure}[!t]
\begin{center}
\epsfig{file=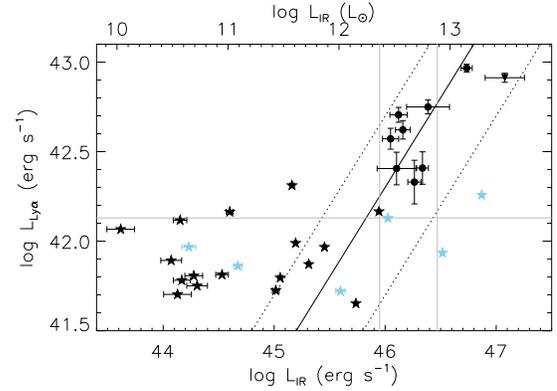,width=8.0cm}
\caption{IR to Ly$\alpha$ luminosities. Filled stars are sources
at $z = 0.3$, filled circles are at $z = 2.3$.
The filled triangle is the only ULIRG detected in the shallow
MIPS survey. Black indicates galaxies, and blue are unclassified LAEs.
The solid diagonal line corresponds to a flux ratio of 0.02~\%, other diagonal lines are 0.05 and 0.005~\%. Grey lines mark the two MIPS detection limits in the deep and wide surveys (vertical), as well as the flux limit in Ly$\alpha$ for the z23 sample (horizontal).}
\label{fig:IRbol}
\end{center}
\vspace{-0.6cm}
\end{figure}

In Fig.~\ref{fig:IRbol} the Ly$\alpha$ luminosities are shown as a function of the infrared luminosities of the galaxies. Here and in the following analysis, we have chosen to be
conservative and have excluded all AGN from the samples.
In Sect.~\ref{sec:ulirgcolour} we return to the question of AGN and
test how robust the results are against AGN inclusion.
In Fig.~\ref{fig:IRbol} the z23 LAE candidates with MIPS detections seem to follow a given
trend between the two flux measurements. The best-fit ratio between
Ly$\alpha$ and infrared luminosity is $\sim 0.02$~\%. 
In the z03 sample, the Ly$\alpha$ luminosity
interestingly stays constant as a function of infrared luminosity.
This indicates that the physical processes governing the Ly$\alpha$
and the IR luminosities are not related at low Ly$\alpha$ and/or IR
luminosities, although the relation seen at bright luminosities is
based on small number statistics. As the blue points in this
case (sources with no previous identification as either galaxy or AGN) are mixed in the population of galaxy LAEs based on Ly$\alpha$ luminosity, these are hereafter considered as normal LAE candidates.

The bolometric luminosity can be calculated from the infrared and ultraviolet luminosity according to
\begin{equation}\label{eq:bol}
L_{bol} = L_{IR}+L_{1600},
\end{equation}
where
\begin{equation}
L_{1600} = \lambda_{1600} \, f_{1600} \, 4 \, \pi \, dL^2.
\end{equation}
This can further be converted to a dust unobscured star formation rate, assuming that the bolometric luminosity includes all the re-processed light from star forming regions (Kennicutt 1998):
\begin{equation}\label{eq:bolsfr}
SFR = 4.5 \times 10^{-44} \times L_{bol} \,  \mathrm{[erg \, s^{-1}]}.
\end{equation}
The star formation rates found from the bolometric luminosity are in the range $500 - 5000$~M$_{\odot}$~yr$^{-1}$ for the non-AGN z23 LAEs and $2 - 3300$~M$_{\odot}$~yr$^{-1}$ for the non-AGN z03 LAEs. Comparing the star formation rates found from the Ly$\alpha$ line and from the bolometric luminosity, we find a median ratio of $0.0043 \pm 0.0025$ and $0.034 \pm 0.18$ for the high and z03 non-AGN LAEs, where the error bars indicate the spread in the values. As the SFR found from the bolometric luminosity is the total SFR of the galaxy, tracing the same population of star forming objects as those creating the Ly$\alpha$ emission, the Ly$\alpha$ escape fraction can be calculated as the ratio between these two values. In the z23 sample, this escape fraction is $\sim 0.4$\% for infrared selected Ly$\alpha$ emitters, whereas it covers the whole range from a few to 100\% in the z03 sample. Furthermore, for the z23 sample, five out of the 9 non-AGN LAEs have SFRs that should make them observable in the \emph{Chandra} X-ray images of the COSMOS field ($\log L_X > 43.3$~erg~s$^{-1}$), if we assume the SFR-X-ray luminosity conversion of Ranalli et al.~(2003). That they are not must indicate that the X-ray sources are obscured by dust.

\begin{figure}[!t]
\begin{center}
\epsfig{file=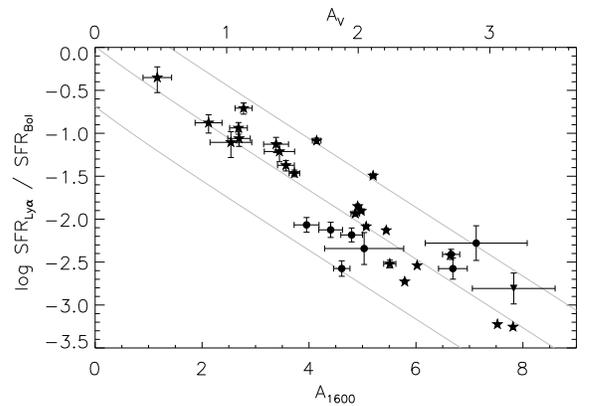,width=8.0cm}
\caption{Star formation rate ratios between Ly$\alpha$ and infrared derived SFRs versus A$_{1600}$. Stars are sources at $z = 0.3$, and filled circles are sources at $z = 2.3$. 
The solid lines follow the relation of constant Ly$\alpha$ restframe equivalent width, with the lower, middle and upper lines representing $EW = 20, 100, 400$~{\AA}.} 
\label{fig:IRsfr}
\end{center}
\vspace{-0.8cm}
\end{figure}

In Meurer et al.~(1999), a relation between the ratio of infrared to ultraviolet flux and the dust attenuation $A_{1600}$ was derived:
\begin{equation}\label{eq:sfrav}
\log \frac{L_{IR}}{L_{1600}} = \log (10^{0.4 \times A_{1600}} - 1)  + 0.076.
\end{equation}
 The spread in $A_{1600}$ for the galaxies here is large. The attenuation in both samples reaches as high as eight magnitudes in $A_{1600}$. In the z03 sample the attenuation spreads down to one magnitude, whereas the z23 galaxies all have $A_{1600} > 3$. The distribution is flat in both samples. 
A plot of the SFR ratios against the dust attenuation derived according to this equation is found in Fig.~\ref{fig:IRsfr}.
The SFR ratios, which are proxies of the Ly$\alpha$ escape fraction, are seen to decrease with increasing dust attenuation, similar to what is seen in local galaxies (Atek et al.~2009). 
The appearance of this plot, however, stems from the relation
between the $x-$ and $y-$axis and Ly$\alpha$ equivalent width (EW).
Lines of constant EW have been overplotted in Fig.~\ref{fig:IRsfr}.
All objects have EWs of roughly $20 - 400$~{\AA}.

\section{ULIRG fraction and redshift evolution}
The finding that Ly$\alpha$ emitting galaxies can also be very IR-bright, hence dusty, is not a novel result. Chapman et al.~(2003, 2005) showed that a set of sub-mm galaxies had Ly$\alpha$ in emission. Very red LAEs have also been found in other surveys. At $z\sim3.1$, Nilsson et al.~(2007; N07, $\log L_{Ly\alpha} > 41.88$~erg~s$^{-1}$) found one red LAE in a sample of 24 and Lai et al.~(2008; L08, $\log L_{Ly\alpha} > 42.07$~erg~s$^{-1}$) presented four red LAEs among 162, also at $z \sim 3.1$. To determine the fraction of ULIRGs in the $z\sim 3.1$ surveys, we studied the ULIRG template SEDs of Vega et al.~(2008) and find that $z=3.1$ ULIRGs must be detected in the \emph{Spitzer} IRAC bands to the flux limits of ECDF-S. Furthermore, the colours in the IRAC bands have to be red; $R-Ch4 \, (8\mu\mathrm{m}) > 2.5, Ch1\, (3.6\mu\mathrm{m}) - Ch4 > 2$. In the N07 sample, only one object is detected in the IRAC bands, with $R-Ch4 = 6.5$, resulting in one ULIRG in this sample (1/24). In the L08 sample, 18 galaxies are detected in the IRAC bands. However, the colours of these objects, including the red objects, are all $Ch1-Ch4 < 2$ (their Fig. 2). As a result, no ULIRGs are found in their survey (0/162). Combining these two samples, the number of ULIRGs at $z\sim 3.1$ are 1 out of 186, resulting in a ULIRG fraction of $0.5^{+1.3}_{-0.5}$\%, where the $1\sigma$ error bars are based on Poisson statistics.  
 Finally, Colbert et al.~(2006) found that three out of 22 $z \sim 2.3$ LAEs ($14^{+13}_{-7}$\%, excluding their detections of ULIRGs in Ly$\alpha$ blobs\footnote{It is unclear what mechanisms power Ly$\alpha$ blobs, and if they are a distinct population of objects or simply the tail of the size distribution of LAEs. Due to these uncertainties, we chose a conservative selection and disregard these sources here.}, see also Francis et al.~2001) were bright in the observed infrared, and that they were ULIRGs. The percentage of ULIRGs
 in the $z = 2.3$ sample presented here is $ 14^{+8}_{-5}$\%, if only the objects in the deep survey are considered (7/50). This number is a lower estimate, as the sensitivity of the deep COSMOS MIPS data only reaches $\log L_{IR} = 12.4 \, \mathrm{L}_{\odot}$ at $z = 2.3$. In the $z = 0.3$ sample, the results are complete in infrared luminosity, and the percentage of ULIRGs in the sample is $20^{+12}_{-8}$\% (6/30), where we again have included 6 unclassified objects but not the AGN. We
return to the robustness also of this result against AGN
inclusion in Sec.~\ref{sec:ulirgcolour} below.
  
 In Fig.~\ref{fig:frac} the ULIRG fractions are plotted as a function of redshift.
\begin{figure}[!t]
\begin{center}
\epsfig{file=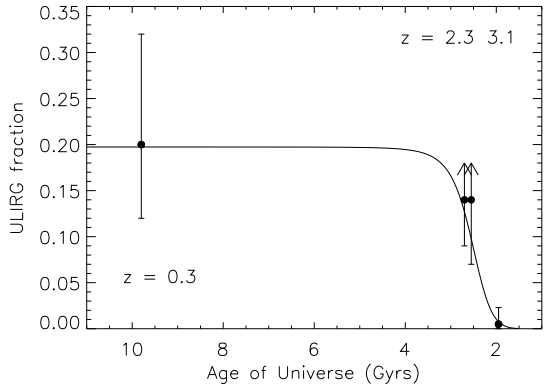,width=8.0cm}
\caption{ULIRG fraction in Ly$\alpha$ samples as a function of age of the Universe. The points at $z \sim 2.3$ are lower limits since both surveys were incomplete in infrared luminosity. A transition from nearly no ULIRGs to $\sim 10-20$\% ULIRGs in Ly$\alpha$ selected samples is seen at $z \sim 2.5$.}
\label{fig:frac}
\end{center}
\vspace{-0.8cm}
\end{figure}
The data points are few, but there is a clear trend for the ULIRG
fraction to grow from high redshifts up to the present day. There
even seems to be a hint that the transition from almost zero ULIRG
fraction to the current value is rather sudden. For the sole
purpose of illustrating the time and steepness of the transition, we
overplot in Fig.~\ref{fig:frac} a hyperbolic tangents function of
the form:
\begin{equation}\label{eq:tanh}
UF(z) = \frac{UF_{0}}{2} (1-\tanh(\theta \,\, (z - z_{tr}))).
\end{equation}
In this equation, $UF$ is the ULIRG fraction, $\theta$ represents
the steepness of the transition, and $z_{tr}$ is the transition redshift.
The function plotted in Fig.~\ref{fig:frac} has $UF_0 = 0.20$
(present-day fraction), $\theta = 2.28$, and $z_{tr} = 2.52$.
This means that the transition redshift for the ULIRG fraction is $z \sim 2.5$; the ULIRG fraction is negligible at higher redshifts, whereas at lower redshifts it approaches $\sim 25$\%.

\subsection{Robustness of the result}\label{sec:ulirgcolour}
\begin{figure}[!t]
\begin{center}
\epsfig{file=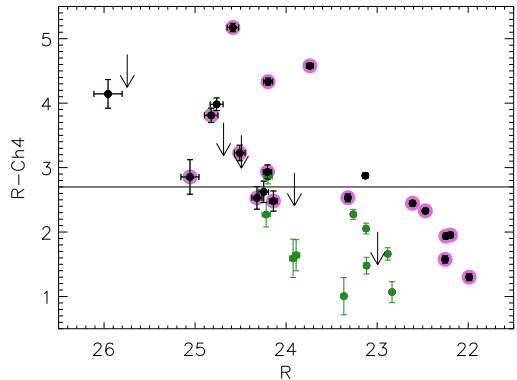,width=8.0cm}
\caption{$R-Ch4$ colours for the LAEs at $z=2.3$ with at least $3\sigma$ detections in each band. For a description of the different data points, see text. The solid line marks our colour selection criteria ($R-Ch4 > 2.7$).}
\label{fig:colours}
\end{center}
\vspace{-0.6cm}
\end{figure}

For the definition of the Ly$\alpha$-selected ULIRG subsamples we
used the common definition based on $L_{IR}$. IR-selected ULIRG
samples are known to have a significant (15-50\%) fraction of
galaxies with AGN components, but it is also known that
their total luminosity is often dominated by star formation rather
than by the AGNs (Veilleux et al.~2009). Nevertheless, to make sure that our
Ly$\alpha$ selection did not bias the results, we chose the most
conservative approach and therefore removed all X-ray detected objects from
the samples. We now ask the question of whether a less conservative
approach is possible, and if that would change the conclusions.
We therefore searched for an additional test to certify that
a candidate is a ULIRG. Using the ULIRG template SEDs of Vega
et al.~(2008),
we find that
a useful index at $z=2.3$ is optical $R$ band minus Ch4 of IRAC. All
Vega templates have $R-Ch4>2.7$, see Fig.~\ref{fig:colours}. 
The points in the plot show all the LAE candidates at $z=2.3$ from Nilsson et al.~(2009) with $> 3\sigma$ detections in $R$ and $Ch4$. The black points are those detected with MIPS, whereas the green points are not MIPS-detected. X-ray detected objects are marked with a pink ring. Four ULIRGs are not detected in $Ch4$ and are shown with upper limits. It can be seen from this plot that the flux selection with MIPS very cleanly samples the region of colour space where ULIRGs are expected to be, except the very brightest AGN that are too blue for this selection. Selecting ULIRGs based on the infrared flux and the $R-Ch4$ colour, and including all objects that have the correct colours within $1\sigma$, changes the $z=2.3$ sample by increasing the number of ULIRGs from seven of a total 50 LAE candidates in the first selection to eight of a total 60 candidates in the second selection. The new fraction is $13^{+7}_{-5}$\%, in perfect agreement with the initial result.

The ULIRG fraction at $z=0.3$ was obtained using the
Deharveng et al.~(2008) sample but excluding the AGN reported by
Cowie et al.~(2009). If the AGN are included, the ULIRG fraction drops to $19^{+12}_{-8}$\% rather than the previous $20^{+12}_{-8}$\%, while
ignoring the Deharveng et al.~(2008) sample. Using only the sample
of Cowie et al.~(2009) gives a ULIRG fraction of $12^{+16}_{-4}$\%. All of those results are
identical to within $1\sigma$ errors, so the $z=0.3$ result
is also robust against details of the sample selection.

\section{Conclusion}
We have here presented the infrared properties of two samples of Ly$\alpha$ emitters, at $z = 0.3$ (24 objects) and $z = 2.3$ (25 objects). The samples were originally Ly$\alpha$-selected but here we considered only the subsamples that were also detected with \emph{Spitzer}.
The two subsamples were selected to determine the fraction of powerful dusty starburst
galaxies (ULIRGs) among LAEs and the redshift evolution of this fraction. Our most important conclusions are the following.

\emph{i)} At all redshifts below $z = 3$, a non-zero fraction of LAEs are found to be ULIRGs. This
result directly contradicts the classic view that dust and Ly$\alpha$ emission are
mutually exclusive, and it holds together with the finding that Ly$\alpha$ emission at high redshifts
correlate with metallicity (M{\o}ller et al.~2004) makes the case that we must re-think the importance of dust for the
Ly$\alpha$ escape fraction. 

\emph{ii)} There is evidence for a strong evolution in ULIRG fraction from redshift 3.1 to the present
universe. The fractions derived have been shown to be very robust against different selection criteria. This explains why there was severe disagreement about the colour of Ly$\alpha$ galaxies
originally. At redshifts around three they were reported to be young, blue, low-dust starbursts (Warren \& M{\o}ller 1996, Fynbo et al.~2000), while
at redshifts closer to two they were reported to be red/dusty (Stiavelli et al.~2001).

\emph{iii)} There may be evidence that the evolution of the ULIRG fraction is not linear in time, but
rather that there is a jump from redshift three to two, and then a more gradual evolution to the present
time. Even though the evidence is merely suggestive at present, this jump does coincide with the
number density evolution of quasars and suggests a connection between those classes of objects.
This question bears strongly on the connection between starbursts and the formation of
quasars, so it should be investigated more thoroughly via dedicated surveys of LAEs at several
different redshifts in the range $z = 2$ to $z = 3.5$. We provide a
simple formalism in the form of a three parameter function which is
well suited to future tests of how sharp the transition is. 

From the accumulation of evidence, it is clear that some Ly$\alpha$ emitting galaxies are very red and dusty, and vice versa that at least some ULIRG galaxies exhibit Ly$\alpha$ emission. 
As was shown in e.g. Finkelstein et al.~(2009), it is possible that for some LAEs the Ly$\alpha$ emission is not affected by the dust at all, or at least not along the line-of-sight towards us. The possibility that Ly$\alpha$ equivalent widths may be enhanced in dusty environments has also been discussed in, e.g., Neufeld~(1991), Hansen \& Oh~(2006), and Finkelstein et al.~(2008, 2009). Interestingly, one of the ULIRG LAEs found at high redshift in the Nilsson et al.~(2009) sample also has by far the largest equivalent width of their sample, with a restframe Ly$\alpha$ equivalent width of nearly $800$~{\AA}.
Considering that samples of LAEs now have for several years consistently included red, dusty galaxies, it is advisable to stop considering Ly$\alpha$ emitters as one type of galaxy, as we have seen in this and previous publications (Nilsson et al. 2009) that Ly$\alpha$ emitters exhibit a range of properties and that this range increases with decreasing redshift.

\begin{acknowledgements}
The authors wish to thank Daniel Schaerer and Anne Verhamme for
useful discussions. We would also like to thank the anonymous referee
for useful comments that have significantly improved the presentation
of our results.
\end{acknowledgements}

\end{document}